\documentclass[aps,prc,twocolumn,10pt, superscriptaddress, showpacs, floatfix]{revtex4-1}
\usepackage{amssymb,epsfig}

\graphicspath{{figures/}}
\begin{document}

\title{Electric dipole response of neutron-rich Calcium isotopes in relativistic quasiparticle time blocking approximation}

\author{Irina A. Egorova}
\affiliation{Department of Physics, Western Michigan University, Kalamazoo, MI 49008, USA}
\affiliation{Bogoliubov Laboratory of Theoretical Physics, JINR, Dubna, 141980, Russia}
\affiliation{FSBI SSC RF ITEP of NCR Kurchatov Institute, Moscow RU-117218, Russia}
\author{Elena Litvinova}
\affiliation{Department of Physics, Western Michigan University, Kalamazoo, MI 49008, USA}
\affiliation{National Superconducting Cyclotron Laboratory, Michigan State University, East Lansing, MI 48824-1321, USA}

\date{\today}

\begin{abstract}

New results for electric dipole strength in the chain of even-even Calcium isotopes with the mass numbers A = 40 -- 54 are presented. Starting from the covariant Lagrangian of Quantum Hadrodynamics, spectra of collective vibrations (phonons) and phonon-nucleon coupling vertices for $J \leq 6$ and normal parity were computed in a self-consistent relativistic quasiparticle random phase approximation (RQRPA). These vibrations coupled to Bogoliubov two-quasiparticle configurations (2q$\otimes$phonon) form the model space for the calculations of the dipole response function in the relativistic quasiparticle time blocking approximation (RQTBA). The results for giant dipole resonance in the latter approach are compared to those obtained in RQRPA and to available data. Evolution of the dipole strength with neutron number is investigated for both high-frequency giant dipole resonance (GDR) and low-lying strength. Development of a pygmy resonant structure on the low-energy shoulder of GDR is traced and analyzed in terms of transition densities. A dependence of the pygmy dipole strength on the isospin asymmetry parameter is extracted.

\end{abstract}


\maketitle


\section{Introduction}
\label{sec:intro}


Nuclear response to various experimental probes is the most informative source of knowledge about the structure of atomic nuclei. As dipole response largely dominates nuclear spectra, it is presently the best known one from the experimental point of view. A large corpus of data has been collected over the decades on various aspects of nuclear dipole response, see a recent review \cite{SavranAumannZilges2013}. However, theoretical description of dipole strength functions still suffer from ambiguities of underlying models. Despite impressive progress of microscopic models based on the density functionals, shell-model or bare nucleon-nucleon interaction, an accurate quantitative description of dipole spectra in medium-mass and heavy nuclei is still a challenge, which is especially actual for the low-energy dipole response because of its relevance for r-process nucleosynthesis \cite{Goriely1998}. In neutron-rich medium-mass nuclei the low-lying dipole strength exhibits a relatively collective resonant character, which has been revealed in a series of theoretical investigations, see review \cite{PaarVretenarKhanEtAl2007}, and, therefore, this part of excitation spectra is often called pygmy dipole resonance (PDR) or soft dipole mode.
From the macroscopic point of view, the pygmy dipole resonance is a collective excitation mode associated with oscillation of a neutron skin against an inert core in neutron-rich nuclei. This picture is also supported by self-consistent microscopic models. The discovery of the pygmy dipole resonance in stable and unstable neutron-rich nuclei and especially its relevance to the astrophysical r-process nucleosynthesis directed the interest towards systematic studies of evolution of the PDR with N/Z ratio, or isospin asymmetry parameter \cite{SavranAumannZilges2013}.

Progress in experimental techniques, such as nuclear resonance fluorescence, has enabled highly accurate measurements of low-lying dipole strength below the nucleon threshold. Such measurements were performed mostly in medium-mass nuclei, in particular, $^{112,116,120,124}$Sn  \cite{GovaertBauwensBryssinckEtAl1998,OezelEndersNeumann-CoselEtAl2007},  $N = 82$ isotones \cite{ZilgesVolzBabilonEtAl2002,VolzTsonevaBabilonEtAl2006}, $^{208}$Pb \cite{RyezayevaHartmannKalmykovEtAl2002}, while some comprehensive studies in light mass region for such nuclei as $^{40,44,48}$Ca \cite{HartmannEndersMohrEtAl2000,HartmannBabilonKamerdzhievEtAl2004} have been also reported. Experiments with radioactive beams allowed measurements of the pygmy dipole resonance in neutron-rich nuclei where it is located above the neutron emission threshold \cite{KlimkiewiczAdrichBoretzkyEtAl2007,RossiAdrichAksouhEtAl2013}. Lately, starting from the pioneering work of D. Savran and co-authors \cite{SavranBabilonBergEtAl2006}, isospin splitting of pygmy dipole resonance was actively studied by applying hadronic probes, such as alpha particles \cite{EndresSavranBergEtAl2009,EndresLitvinovaSavranEtAl2010,EndresSavranButlerEtAl2012,DeryaSavranEndresEtAl2014}, protons \cite{TamiiPoltoratskaNeumann-CoselEtAl2011,HashimotoKrumbholzReinhardEtAl2015} and heavier nuclei \cite{PellegriBraccoCrespiEtAl2014,KrzysiekKmiecikMajEtAl2016} complementary to photon scattering. 
Systematic theoretical and experimental studies of the pygmy dipole resonance lead to important conclusions about nuclear symmetry energy and dipole polarizability \cite{Roca-MazaVinasCentellesEtAl2015,ReinhardNazarewicz2010}.

In this work we focus on studies of the electromagnetic dipole strength in even-even Calcium isotopes with neutron excess. Although neutron-rich Calcium isotopes do not lie on the r-process nucleosynthesis path and, therefore, are not in the direct relevance to the synthesis of heavy elements, experimental and theoretical investigations of dipole response in a long isotopic chain of Calcium can provide invaluable information on nuclear symmetry energy and on the equation of state. In addition, for the approaches based on the density functional Calcium mass region is of special interest because it is located in the proximity of the low-mass limit of model applicability. 
 
The giant dipole resonance in Calcium isotopes is known experimentally since relatively long ago. Absolute nuclear photoabsorption cross section for electric dipole excitations above 10 MeV has been measured for $^{40}$Ca in Ref.  \cite{AhrensBorchertCzockEtAl1975}. Measurements for $^{48}$Ca  were performed in Ref. \cite{OKeefeThompsonAssafiriEtAl1987}.
Later, some evaluated data for photoabsorption cross section in the same energy range became available for the chain of even-even Calcium isotopes $^{40,42,44,48}$Ca in \cite{erokhova2003giant4421225}. 

The new generation of the radioactive beam facilities allows for nuclear structure data on heavier Calcium isotopes far from stability. Already  long ago, an evidence for a new shell gap at neutron number N = 32 was reported for $^{52}$Ca  \cite{HuckKlotzKnipperEtAl1985} as a result of measurements of the two-proton knockout reaction at ISOLDE/CERN.  
On the other hand, it was observed recently in Radioactive Isotope Beam Factory at RIKEN \cite{SteppenbeckTakeuchiAoiEtAl2013} that the shell structure of $^{54}$Ca highlights a sizable subshell closure at neutron number N = 34 in isotopes far from stability. Moreover, heavier calcium isotopes up to $^{60}$Ca can be produced and investigated at experimental radioactive beam facilities, such as National Superconducting Cyclotron Laboratory (NSCL) and RIKEN \cite{TarasovMorrisseyAmthorEtAl2013}, due to the increase of primary beam intensities, new beam development and advances in experimental techniques. Neutron-rich Calcium isotopes continue to attract substantial interest since they show shell evolution with new magic numbers at N = 32, 34 and serve as a very convenient test case for theoretical models €™\cite{GadeSherrill2016}, stimulating new developments on the theory side, in particular, €˜in the so-called 'ab initio' framework \cite{WienholtzBeckBlaumEtAl2013}, which is able to reproduce masses and low-energy spectroscopy of light nuclei.

High-frequency giant collective excitations in the Calcium mass region are also becoming accessible by theoretical methods employing chiral nucleon-nucleon interactions \cite{Bacca2014}. As complex many-body correlations are very important for the formation of giant resonances and for their low-energy fine features, a variety of self-consistent approaches beyond random phase approximation (RPA) were developed and applied to Calcium. Nuclear field theory \cite{BesBrogliaDusselEtAl1976,BortignonBrogliaBesEtAl1977,BertschBortignonBroglia1983,MahauxBortignonBrogliaEtAl1985,ColoBortignon2001}, quasiparticle-phonon model \cite{Soloviev1992} second RPA (SRPA) \cite{ShibuyaMcKoy1970,TakayanagiShimizuArima1988,YannouleasJangChomaz1985,PapakonstantinouRoth2009,GambacurtaGrassoCatara2011}, mode coupling model \cite{Schuck1976} and method of chronological decoupling of diagrams (later renamed to time-blocking approximation) \cite{Tselyaev1989,KamerdzhievTertychnyiTselyaev1997,Tselyaev2007,LitvinovaTselyaev2007,LitvinovaRingTselyaev2007,TertychnyTselyaevKamerdzhievEtAl2007,LyutorovichTselyaevSpethEtAl2015} account for two-particle-two-hole excitations with various coupling schemes, which are responsible for damping effects of both giant resonance and low-energy strength. Continuum effects are known to contribute to the damping of giant resonances and to cause an additional broadening \cite{ShlomoBertsch1975,KhanSandulescuGrassoEtAl2002,KamerdzhievSpethTertychny2004,LitvinovaTselyaev2007,DaoutidisRing2011}. Most of the studies of collective excitations in Calcium mass region in the approaches beyond RPA are, however, limited by closed-shell systems, with only a few exceptions \cite{HartmannBabilonKamerdzhievEtAl2004,TertychnyTselyaevKamerdzhievEtAl2007,KnappLoIudiceVesely2015}. 

This work presents a systematic investigation of electric dipole spectra of even-even Calcium isotopes with A = 40 -- 54 within the relativistic quasiparticle time blocking approximation (RQTBA) developed originally in Ref. \cite{LitvinovaRingTselyaev2008}. The method is based on the relativistic meson-exchange nuclear Lagrangian of Quantum Hadrodynamics and extends the covariant response theory by effects of retardation. The retardation, or time-dependence of meson-exchange interaction, is neglected in the covariant density functional theory \cite{Ring1996,VretenarAfanasjevLalazissisEtAl2005} and in the approaches on the level of two-quasiparticle configurations, such as relativistic quasiparticle random-phase approximation (RQRPA) \cite{PaarRingNiksicEtAl2003}. It is restored in RQTBA in an approximate way taking into account the most important  (resonant) effects of temporal non-localities, essential at the relevant excitation energies ($\sim$ 0-50 MeV), on the equal footing with superfluid pairing.  In the original version of RQTBA they are modeled by coupling of quasiparticles to collective vibrations within 2q$\otimes$phonon coupling scheme, and in the extended versions within phonon-phonon coupling \cite{LitvinovaRingTselyaev2010,LitvinovaRingTselyaev2013} and 
2q$\otimes$Nphonon \cite{Litvinova2015} configurations. The method was applied with great success to dipole spectra of various closed and open-shell medium-mass nuclei \cite{LitvinovaRingTselyaev2008,LitvinovaRingTselyaevEtAl2009,MassarczykSchwengnerDoenauEtAl2012}. In particular, it has described the observed spin-isospin splitting of pygmy dipole resonance \cite{EndresLitvinovaSavranEtAl2010,EndresSavranButlerEtAl2012,LanzaVitturiLitvinovaEtAl2014}, isoscalar dipole modes \cite{PellegriBraccoCrespiEtAl2014,KrzysiekKmiecikMajEtAl2016}, and stellar reaction rates of r-process nucleosynthesis \cite{LitvinovaLoensLangankeEtAl2009}. Recently, an extension of the R(Q)TBA to spin-isospin excitations has become available
\cite{MarketinLitvinovaVretenarEtAl2012,LitvinovaBrownFangEtAl2014,RobinLitvinova2016}. 

The structure of the article is as follows: after a brief review of the formalism in Section \ref{sec:thmod},  in Section \ref{sec:disc} we discuss the evolution of the electric dipole strength in the excitation energy region 0-30 MeV with neutron number, in comparison with available data, and the development of the soft mode at low energies. Then, by a careful analysis of the transition densities, the location of the pygmy dipole resonance is determined and its systematics with respect to the asymmetry parameter is extracted. Finally, Section \ref{summary} presents conclusions and outlook.


\section{Formalism}
\label{sec:thmod}

In this Section, we give a brief survey of the relativistic quasiparticle time blocking approximation in the neutral particle-hole channel. It is conveniently  formulated in terms of nuclear response function in the basis of states of Dirac-Hartree-Bogoliubov quasiparticles $\{k_i,\eta_i\}$ \cite{KucharekRing1991}, where the indices $k_i$ run over the complete sets of the single-quasiparticle quantum numbers including states in the Dirac sea and indices $\eta_i$, numerate forward (+) and backward (-) components of the matrix elements in the Nambu space.  In practice, this basis is generated by a self-consistent solution of the relativistic mean-field problem \cite{Ring1996,VretenarAfanasjevLalazissisEtAl2005} with superfluid pairing correlations, which forms the content of the Covariant Density Functional Theory (CDFT). The final equation for the nuclear response function $R(\omega)$ in the resonant quasiparticle time blocking approximation is given by the following equation:
\begin{eqnarray}
R_{k_{1}k_{4},k_{2}k_{3}}^{\eta\eta^{\prime}}(\omega)&=\tilde{R}_{k_{1}k_{2}%
}^{(0)\eta}(\omega)&\delta_{k_{1}k_{3}}\delta_{k_{2}k_{4}}\delta_{\eta
\eta^{\prime}} \nonumber \\
&+ \tilde{R}_{k_{1}k_{2}}^{(0)\eta}(\omega)&\sum\limits_{k_{5}%
k_{6} \eta^{\prime\prime}}{\bar{W}}_{k_{1}k_{6}%
,k_{2}k_{5}}^{\eta\eta^{\prime\prime}}(\omega)R_{k_{5}k_{4},k_{6}k_{3}}%
^{\eta^{\prime\prime}\eta^{\prime}}(\omega), \nonumber \\
&&
\label{respdir}
\end{eqnarray}
which is the spectral representation of the Bethe-Salpeter equation (BSE) projected onto the particle-hole channel \cite{Tselyaev2007, LitvinovaTselyaev2007,LitvinovaRingTselyaev2008}. The latter implies that the indices $\eta,\eta^{\prime}$ are combined pair indices, so that $\eta = + = \{\eta_1,\eta_2\} = \{+,-\}$ and $\eta = - = \{\eta_1,\eta_2\} = \{-,+\}$.  $\tilde{R}^{(0)}(\omega)$ is the propagation of two quasiparticles in the mean field between acts of interaction with the following amplitude:
\begin{eqnarray}
{\bar{W}}_{k_{1}k_{4},k_{2}k_{3}}^{\eta\eta^{\prime}}(\omega)&=&\tilde{V}%
_{k_{1}k_{4},k_{2}k_{3}}^{\eta\eta^{\prime}} \nonumber \\
&&+\Bigl(\Phi_{k_{1}k_{4},k_{2}%
k_{3}}^{\eta}(\omega)-\Phi_{k_{1}k_{4},k_{2}k_{3}}^{\eta}(0)\Bigr)\delta
_{\eta\eta^{\prime}}.\nonumber \\
&& 
\label{W-omega}%
\end{eqnarray}
Here $\tilde V$ is the meson-exchange interaction of the  CDFT with the parameter set NL3$^*$ \cite{LalazissisKaratzikosFossionEtAl2009}
with a non-linear self-coupling of the scalar $\sigma$-meson:
\begin{equation}
{\tilde V} = {\tilde V}^{nl}_{\sigma} + {\tilde V}_{\omega} + {\tilde V}_{\rho} + {\tilde V}_e, 
\label{mexch}
\end{equation}
where retardation effects are neglected in the $\sigma$-, $\omega$-, $\rho$-meson and photon propagators.
This approximation is well justified 
for a single meson-exchange interaction, because the meson masses are large compared to the typical
energy scale in nuclear structure physics. However, when the large attractive scalar field gets strongly
compensated by the vector field, the resulting temporal non-locality, appearing in the spectral representation as an 
energy dependence, can become significant. The latter effects can be accurately taken into account in an approximate way 
considering such a non-locality as a superposition of small-amplitude collective vibrations of nucleonic density with various multipoles.
The underlying mechanism of these vibrations is the meson exchange significantly modified by the nuclear medium. In the self-consistent theory 
the vibrations are generated by the same effective interaction $\tilde V$. They provide a feedback on the single-quasiparticle degrees of freedom acting as mediators of an additional boson-exchange (commonly called phonon-exchange) interaction.  Accounting for these contributions is the main content of the RQTBA. Because of the presence of vibrational solutions at very low energies ($\sim$ 1 MeV), the retardation effects caused by  the exchange of these vibrations become very important. In Eq. (\ref{W-omega}) they are encapsulated in the particle-hole ($\eta = +$) and hole-particle ($\eta = -$) amplitude $\Phi^{\eta}(\omega)$, which, in the resonant time blocking approximation, has the following structure:
\begin{eqnarray}%
\Phi_{k_{1}k_{4},k_{2}k_{3}}^{\eta}(\omega)  = \nonumber \\
= \sum\limits_{\mu\xi}\delta_{\eta\xi}
\Bigl[\delta_{k_{1}k_{3}}\sum\limits_{k_{6}} \frac{\gamma_{\mu;k_{6}k_{2}%
}^{\eta;-\xi} \gamma_{\mu;k_{6}k_{4}}^{\eta;-\xi\ast}}{\eta\omega-E_{k_{1}}-E_{k_{6}%
}-\Omega_{\mu}} + \nonumber \\
+ \delta_{k_{2}k_{4}}\sum\limits_{k_{5}}\frac{\gamma
_{\mu;k_{1}k_{5}}^{\eta;\xi}
\gamma_{\mu;k_{3}k_{5}}^{\eta;\xi\ast}}{\eta\omega-
E_{k_{5}} - E_{k_{2}} - \Omega_{\mu}}\nonumber\\
-\Bigl(\frac{\gamma_{\mu;k_{1}k_{3}}^{\eta;\xi} \gamma_{\mu
;k_{2}k_{4}}^{\eta;-\xi\ast}}{\eta\omega- E_{k_{3}}- E_{k_{2}} -
\Omega_{\mu}} +
\frac{\gamma_{\mu;k_{3}k_{1}}^{\eta;\xi\ast}\gamma_{\mu;k_{4}k_{2}}^{\eta;
-\xi}} {\eta\omega- E_{k_{1}} - E_{k_{4}} -
\Omega_{\mu}}\Bigr)\Bigr],\nonumber
\\
\label{phiphc0}%
\end{eqnarray} 
where we denote the matrix elements of the quasiparticle-phonon coupling vertices as $\gamma_{\mu;k_{1}k_{2}}^{\eta;\xi} =
\gamma_{\mu;k_{1}k_{2}}^{\eta;\xi\xi}$. In Eq. (\ref{phiphc0}) the index $\xi$ stands for the particle-particle ($\xi = +$) and hole-hole ($\xi = -$) components
of the vertex matrix elements computed within RQRPA as well as their frequencies $\Omega_{\mu}$. The index $\mu$ numerates phonon quantum numbers and $E_{k_{i}}$ are the energies of Dirac-Hartree-Bogoliubov quasiparticles. Conventionally, phonon modes with natural parities and multipolarities $J_{\mu} \leq 6$  are included in the RQTBA model space, if their reduced transition probabilities are 5\% or larger than the maximal one for the given multipolarity. The component structure of the matrix elements of the meson-exchange interaction (\ref{mexch}) is given in Ref. \cite{LitvinovaRingTselyaev2008}.

The physical content of the time blocking approximation is that, due to the time projection in the integral part of the BSE, the
two-quasiparticle propagation through states of a more complicated structure than $2q\otimes$ phonon is blocked~\cite{Tselyaev2007}. The 
amplitude $\Phi(\omega)$ in Eq. (\ref{W-omega})  is corrected by the subtraction of itself at zero frequency, in
order to avoid double counting of the quasiparticle-phonon coupling effects, which are implicitly included in the CDFT parameters adjusted to experimental masses and radii of characteristic nuclei \cite{VretenarAfanasjevLalazissisEtAl2005}. 

The BSE (\ref{respdir}) for the response function is solved in both Dirac-Hartree-Bogoliubov and momentum-channel representations \cite{LitvinovaRingTselyaev2008}. The response to a weak external field $P$ is determined by the microscopic strength function $S(E)$:
\begin{eqnarray}
S(E) &=& -\frac{1}{2\pi}\lim\limits_{\Delta\rightarrow+0}Im\
\sum\limits_{k_{1}k_{2}k_{3}k_{4}}\sum\limits_{\eta\eta^{\prime}}
P_{k_{1}k_{2}}^{\eta\ast} \nonumber \\
&& \times R_{k_{1}k_{4},k_{2}k_{3}}^{\eta\eta^{\prime}}
(E + i\Delta)P_{k_{3}k_{4}}^{\eta^{\prime}}.
\label{strf}%
\end{eqnarray}
In this work the operator P is the electromagnetic dipole field:
\begin{equation}
P_{1M} = \frac{eN}{A}\sum\limits_{i=1}^{Z}r_iY_{1M}(\Omega_i) - \frac{eZ}{A}\sum\limits_{i=1}^{N}r_iY_{1M}(\Omega_i).
\end{equation}
The total dipole photoabsorption cross section $\sigma_{E1}(E)$ is related to the strength function $S_{E1}(E)$ as follows:
\begin{equation}
 \sigma_{E1}(E) = \frac{16 \pi^3 e^2}{9{\hbar}c} E S_{E1}(E)\,.
\label{sgth}
\end{equation}
A finite imaginary part $\Delta$ of the energy variable is used in the calculations in order to obtain a smooth envelope of the spectrum. This parameter brings an additional artificial width for each excited state. This width models effectively effects which are not taken into account explicitly in our approach, such as higher-order correlations and escape of particles to continuum. More details about the formalism can be found in Ref. \cite{LitvinovaRingTselyaev2008}.

\begin{figure*}
\begin{center}
\includegraphics*[width=0.86\textwidth]{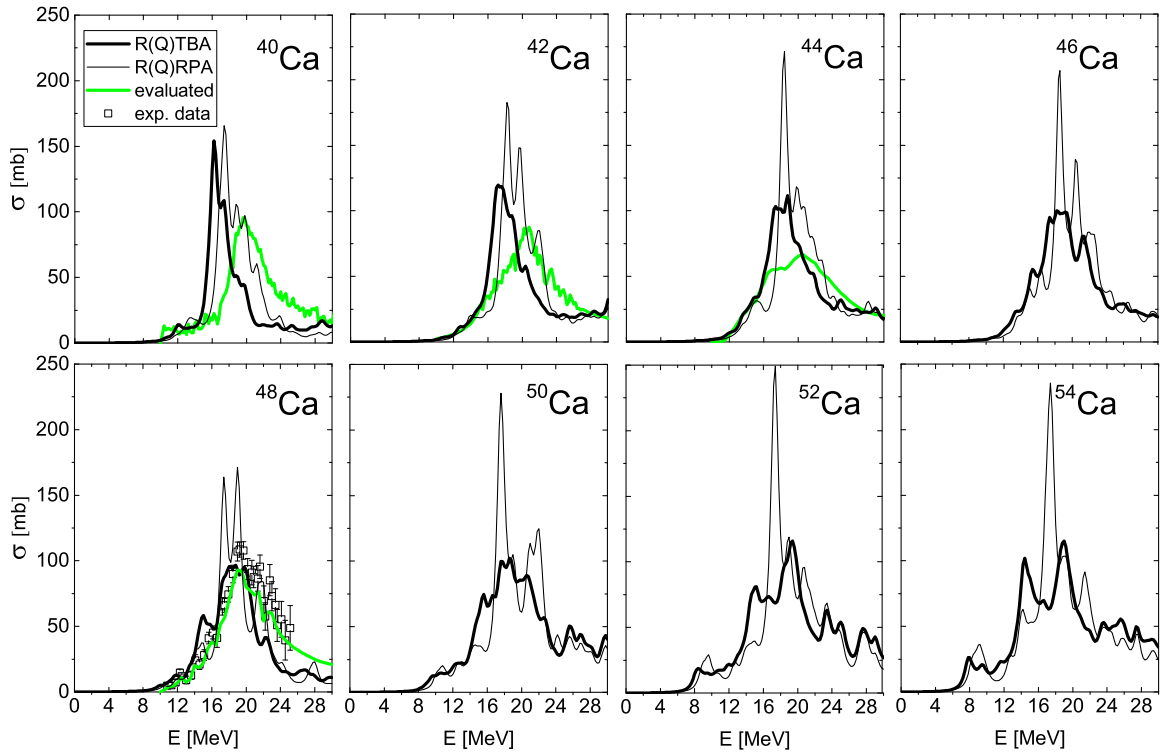}
\end{center}
\caption{Excitation spectra for the Calcium isotopic chain obtained within RQTBA (thick black curves) in comparison to RQRPA  (thin black curves). Evaluated experimental data \cite{EXFOR} are given by green solid curves and the original data \cite{OKeefeThompsonAssafiriEtAl1987} are displayed by the error bars.}
\label{fig:gdrstrength}
\end{figure*}


\section{Results and discussion}
\label{sec:disc}
Calculations within the framework described above were performed for even-even Calcium isotopes with mass numbers A = 40 -- 54 in both RQTBA and RQRPA approaches. The results for dipole photoabsorption cross sections calculated with $\Delta$ = 500 keV in the broad energy region up to 30 MeV are shown in Fig. \ref{fig:gdrstrength} in comparison to experimental data. The first general observation is that in the most of the cases a strong fragmentation of the GDR occurs in RQTBA, as compared to RQRPA. This effect is noticeably stronger in open-shell isotopes, that is in agreement with the overall situation in medium-mass nuclei calculated previously \cite{LitvinovaRingTselyaev2008,MassarczykSchwengnerDoenauEtAl2012}. The clear reason for this is the location of the first collective phonons of the lowest multipolarities, which is always at lower energies in open-shell nuclei than in the closed-shell ones. In addition, as a rule, the transition probabilities for these phonons, which are in the direct relation to their coupling amplitudes $\gamma_{\mu}$, are higher in open-shell nuclei. Besides that, the case of $^{40}$Ca is special for R(Q)TBA in its standard self-consistent formulation, which implies calculations of the phonon spectra $\{\Omega_{\mu}, \gamma_{\mu}\}$ within R(Q)RPA. As a rule, for medium-mass and heavy nuclei and for open-shell light nuclei the lowest phonons are well reproduced by R(Q)RPA. However, in the cases of $^{40}$Ca and $^{16}$O the lowest observed phonons with positive natural parity, such as $2^{+}, 4^{+}, 6^{+}$, have more complicated nature  than one-particle-one-hole and, therefore, are not reproduced in R(Q)RPA. In fact, in R(Q)RPA calculations for $^{40}$Ca the lowest phonons of this kind are above 10 MeV and, therefore, their contribution to the amplitude $\Phi(\omega)$  is very minor, as it can be seen from its pole structure in Eq. (\ref{phiphc0}).  In order to adopt RQTBA to calculations for $^{40}$Ca and $^{16}$O, one has to apply a higher-order approach to the  $2^{+}, 4^{+}, 6^{+}$ phonons, which is not done in the present work. So, the RQTBA result for $^{40}$Ca should be understood with this reservation.  

In $^{42}$Ca the situation is improving: the presence of pairing correlations brings the most important $2^+$ phonon in a better agreement to data although its energy is still by factor 2 larger that the experimental one. Nevertheless, the fragmentation is stronger, and the width is closer to the observed one. The energy of the $2^+$ phonon is very sensitive to the value of the pairing gap which is fine tuned in the present approach. For the calculations shown in Fig. \ref{fig:gdrstrength} pairing gap equal to 2.0 MeV is used, that is close to the empirical value $\Delta_{emp} = 12/\sqrt{A}$ MeV. Alternative calculations were done with pairing gaps reproducing the energy of the first quadrupole state, however, the overall GDR picture is not sensitive to this difference. 

One can see from Fig. \ref{fig:gdrstrength} that the centroids of the GDR in $^{40}$Ca and in its closest neighbor $^{42}$Ca are $\sim$~2~MeV lower than the ones observed in experiments. Interestingly, in other doubly-magic nuclei $^{208}$Pb and $^{132}$Sn, where R(Q)TBA calculations were performed \cite{LitvinovaRingTselyaev2007}, the GDR centroids are also 1.0 -- 1.5 MeV lower than the experimental ones, while in open-shell nuclei the GDR centroids are in a perfect agreement to data \cite{LitvinovaRingTselyaev2008}. This discrepancy appears already in relativistic RPA as well as in RPA based on many Skyrme functionals, and the inclusion of particle-vibration coupling does not change the situation as it preserves the centroids. The reason for this discrepancy and its possible improvement are investigated now.

\begin{figure}
\begin{center}
\includegraphics[width=0.65\textwidth]{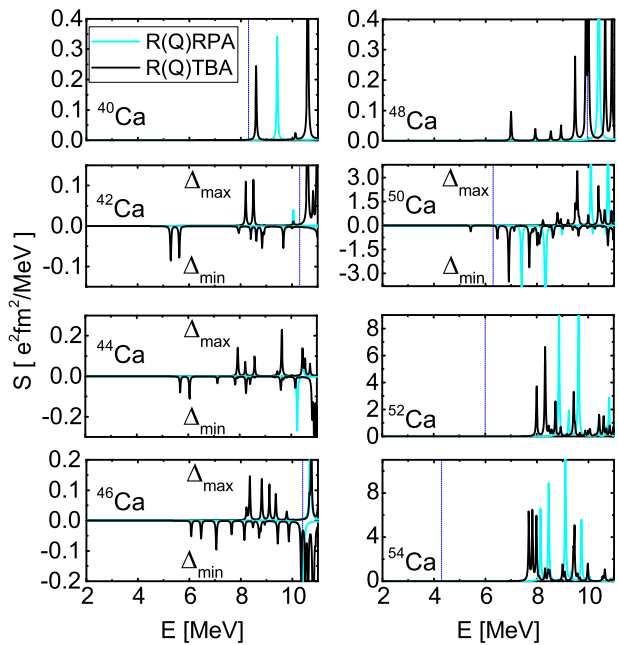}
\end{center}
\caption{Electric dipole strength distribution in Ca isotopes at low energies, calculated in the R(Q)RPA and R(Q)TBA models. The blue vertical lines indicate the experimental particle emission thresholds (proton for $^{40,42}$Ca and neutron for the rest of the isotopes). See text for further details.
}
\label{fig:Pigmystrength}
\end{figure}

Giant dipole resonance in $^{44}$Ca is described relatively well by RQTBA, although its width is still somewhat underestimated. We have not found data for $^{46}$Ca, but in $^{48}$Ca R(Q)TBA calculations provide a very good description. Notice, that the low-lying phonons are reproduced also very well in this nucleus. Overall, the fragmentation of the GDR increases with neutron number and, starting from $^{48}$Ca, it can be seen that a soft mode starts to develop on the low energy shoulder of GDR. This low-energy part of the calculated dipole strength distributions can be seen in more detail in Fig. \ref{fig:Pigmystrength}, where it is shown with the smearing parameter 20 keV. In the open-shell nuclei $^{42,44,46,50}$Ca the spectra are calculated and presented with two values of the pairing gap: $\Delta_{max}$ = 2.0 MeV and $\Delta_{min}$ which reproduces the energy of the first $2^+$ state in each isotope. For $^{52,54}$Ca the low-lying dipole strength is computed only with $\Delta_{max}$ because with smaller paring gap RQRPA solutions show instabilities. The first observation from this Figure is that the fine structure of the low-lying strength is very sensitive to the pairing gap value: retaining similar patterns, the spectra are shifted with respect to each other by $\sim$~2 -- 3 MeV.
Presently, there is no quantitatively satisfactory relativistic approach to pairing correlations in finite nuclei, although the formalism is developed very clearly for the meson-exchange interactions and applied to nuclear matter \cite{KucharekRing1991,SerraRummelRing2001}. In practice, either monopole or finite-range pairing force is employed \cite{VretenarAfanasjevLalazissisEtAl2005}. A separable representation of zero-range pairing force is also available \cite{TianMaRing2009}.
In this work we use the monopole force with the strength which is fine tuned in two ways described above, in order to see the sensitivity of the dipole strength to the pairing gap.
 
Returning to Fig. \ref{fig:Pigmystrength}, one can see also various possibilities for the location of the low-energy dipole strength with respect to the particle emission threshold (shown by dashed blue lines). For instance, in $^{40}$Ca there is no dipole strength below the (proton) emission threshold; in $^{42}$Ca some dipole states appear in this region, due to unblocking effects of pairing correlations, while the amount of the strength depends on  the pairing gap value and, in addition, the proton emission threshold moves up by $\sim$2 MeV being still below the neutron threshold. In the next isotopes $^{44,46,48}$Ca approximately the same amount of strength is seen below 10 MeV while the neutron threshold is moving below the proton one down to 11.13 MeV in $^{44}$Ca and then further down to 9.95 MeV in $^{48}$Ca. Beyond $^{48}$Ca neutrons occupy the next major shell, and the neutron threshold moves further down, so that almost no strength below the threshold is seen in the RQTBA framework.

\begin{figure}
\begin{center}
\includegraphics[scale=0.45]{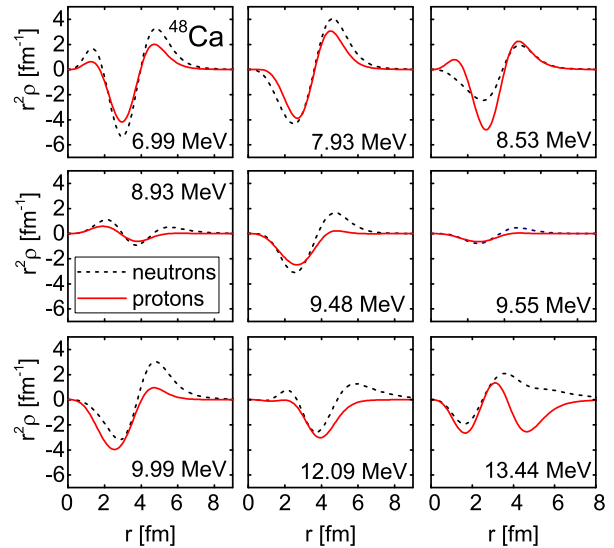}
\end{center}
\caption{RQTBA transition densities for the most prominent peaks in $^{48}$Ca.}
\label{fig:density2}
\end{figure}

As it is already mentioned, systematics of the pygmy dipole resonance as a neutron skin oscillation mode and the related electric dipole polarizability play an important role in constraining the nuclear equation of state \cite{SavranAumannZilges2013,Roca-MazaVinasCentellesEtAl2015}. Presently, the problem with the PDR systematics is that its experimental determination is limited by the particle emission threshold. With only few exceptions \cite{TamiiPoltoratskaNeumann-CoselEtAl2011,HashimotoKrumbholzReinhardEtAl2015}, the measurements have been performed either below or above the threshold with different probes, while the strength in $\sim$1-2 MeV energy window around the threshold remains very uncertain. Besides that,  it is not clear how to separate PDR from GDR in experiments. This is a non-trivial task for the theory as well, because neither GDR nor PDR are single excitation modes. Instead, each of them is fragmented and broadened due to Landau damping (occurring already on the R(Q)RPA level), damping to complex configurations (essentially taken into account in R(Q)TBA) and particle escape to continuum \cite{LitvinovaTselyaev2007}.
However, theoretical methods have a very convenient tool for an approximate separation of these two excitation modes, namely the proton and neutron transitions densities, which reveal a specific pattern characterizing neutron skin oscillation \cite{VretenarPaarRingEtAl2001}: in-phase oscillation of proton and neutron transition densities inside the nucleus with a visible dominance (correlated with the neutron excess) of the neutron component in the surface area. In contrast, GDR shows out of phase proton and neutron density oscillations. Besides that, there is a 'transitional' energy region between PDR and GDR where none of those clear patterns is observed \cite{LitvinovaRingTselyaevEtAl2009}. Thus, in the present work, before making a systematics of the pygmy strength, we have also analyzed the transition densities of the calculated states to establish and approximate border line between PDR and GDR. 
 
Figure \ref{fig:density2} shows proton and neutron transition densities of the most prominent peaks in $^{48}$Ca between 6.99 and 13.44 MeV. Their common feature is nearly in-phase oscillations of proton and neutron density distributions with more or less visible dominance of the neutron component in the surface area, which is the well-established PDR pattern, while the last state at 13.44 MeV starts to deviate from this picture toward the out-of phase behavior inherent for the GDR.   
Although we see stronger fragmentation away from closed shells, the evolution of the transition densities look very similar in all analyzed Calcium isotopes, so that we have set 13.0 MeV as the energy of the separation between PDR and GDR for the whole isotopic chain. 

\begin{figure}
\begin{center}
\includegraphics[width=0.53\textwidth]{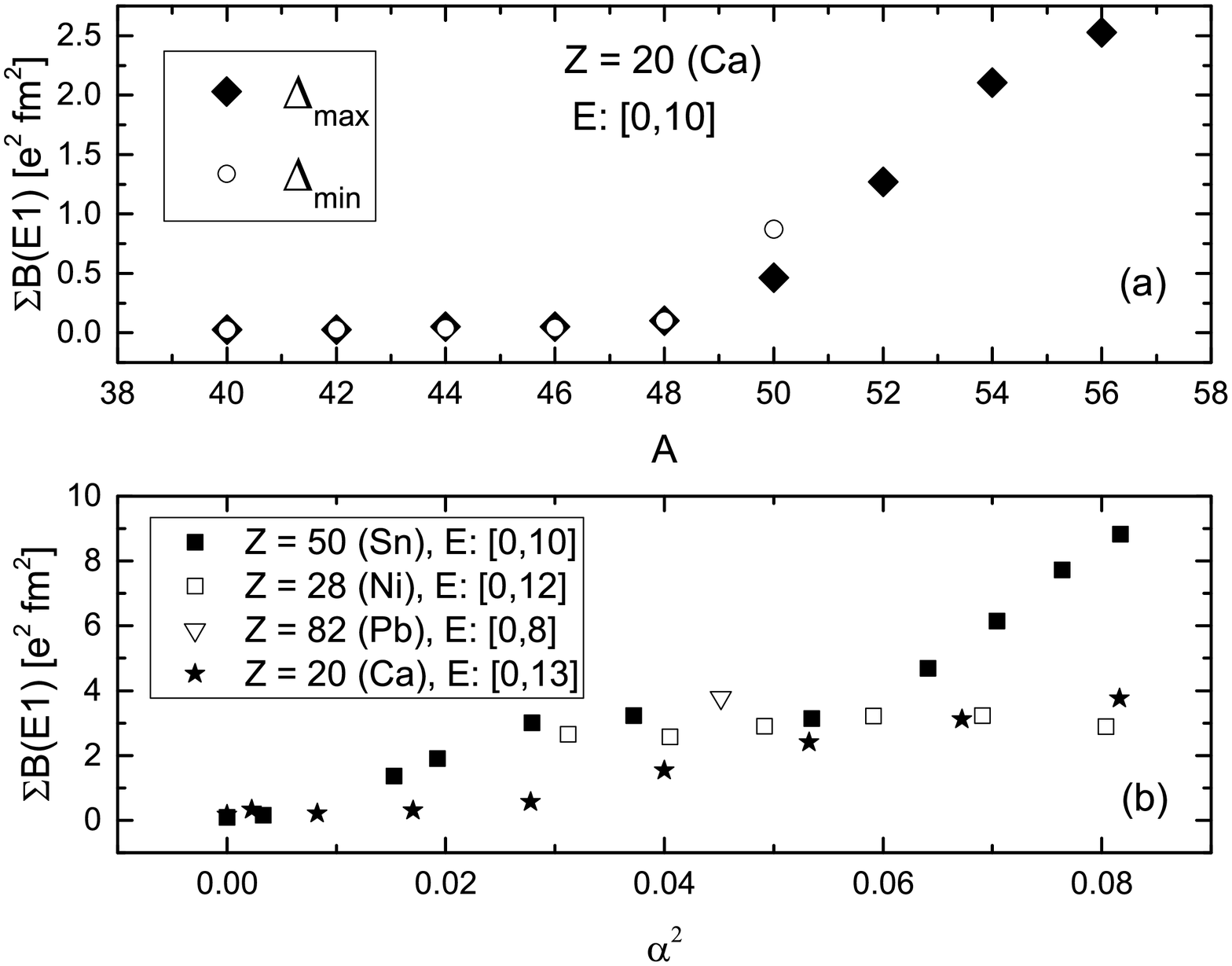}
\end{center}
\vspace{-0.7cm}
\caption{Upper panel (a): Integrated dipole strength below 10 MeV calculated within the RQTBA for Calcium isotopes. Lower panel (b): RQTBA pygmy dipole strength in Ca isotopes, compared to that in Pb, Sn and Ni isotopes \cite{LitvinovaRingTselyaevEtAl2009}, versus squared isospin asymmetry parameter. Energy intervals are shown in square brackets.}
\label{fig:intstr}
\end{figure}

This separation allows for the inclusion of Calcium isotopic chain in the systematics extracted previously from the calculations for Pb, Sn and Ni isotopes \cite{LitvinovaRingTselyaevEtAl2009}, which is shown in the lower panel of Figure  \ref{fig:intstr}. The total pygmy strength is plotted versus the squared isospin asymmetry parameter $\alpha = (N-Z)/A$ for these isotopic chains, while the energy intervals defining the pygmy strength are adjusted by the analyses of proton and neutron transition densities, so that only the states with the typical "pygmy" underlying structure are included in these intervals. The particular values of the upper energy limits are shown in the legend. As discussed in Ref. \cite{LitvinovaRingTselyaevEtAl2009}, the nearly constant values of the pygmy strengths in Sn isotopes for $0.03 \leq \alpha^2 \leq 0.06$ and in Ni isotopes for $0.03 \leq \alpha^2 \leq 0.08$ are caused by the presence of the intruder states $1h_{11/2}$ and $1g_{9/2}$, respectively: neutrons added to these orbits do not induce new low-energy dipole transitions because the next shell has the same parity. In the case of Calcium,  the orbital $1f_{7/2}$ is getting occupied by neutrons when going from N=20 to N=28 ($0.0 \leq \alpha^2 \leq 0.027$), and the latter interval of the isospin asymmetry  parameter also shows a stagnation of the total pygmy strength, while with the opening of a new shell the total pygmy strength increases nearly linearly. For $0.04 \leq \alpha^2 \leq 0.06$ the pygmy strength has very similar values in all isotopic chains under consideration, that points out to the existence of a universal correlation between the isospin asymmetry and the magnitude of the neutron skin oscillation. 

\begin{figure}
\begin{center}
\includegraphics[width=0.50\textwidth]{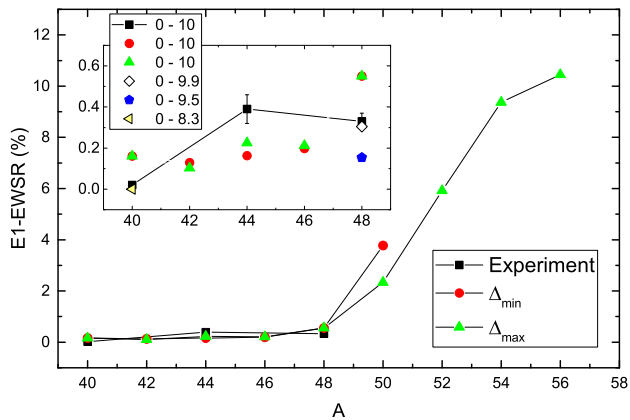}
\end{center}
\caption{Fraction of the energy-wieghted sum rule (EWSR) in Ca isotopes below 10 MeV, relative to Thomas-Reiche-Kuhn sum rule. Inset: EWSR for $^{40-48}$Ca on a smaller scale, compared to data of Ref. \cite{HartmannBabilonKamerdzhievEtAl2004}. For the cases, where there are strong peaks around 10 MeV, EWSR in reduced energy intervals excluding these peaks are presented.}
\label{fig:E1-EWSR-A}
\end{figure}

The upper panel of Fig. \ref{fig:intstr} shows the total pygmy strength in the studied Calcium isotopes below 10 MeV. This is the upper limit of the energy interval which is often investigated experimentally. Here we illustrate the sensitivity of the pygmy strength to the value of the pairing gap. The isotopes $^{42,44,46}$Ca show almost no sensitivity, however, in $^{50}$Ca the pygmy strength calculated with $\Delta_{min}$ gap is by almost factor two larger than in the calculation with $\Delta_{max}$. As mentioned above, calculations for  $^{52,53,56}$Ca in a spherical basis are only possible with the larger pairing gap.

Figure \ref{fig:E1-EWSR-A} displays the fraction of the energy weighted sum rule (EWSR) for the  dipole strength below 10 MeV relative to the Thomas-Reiche-Kuhn (TRK) model independent sum rule as a function of mass number. This quantity shows the behavior, which is similar to the non-weighted strength. The inset gives comparison of the calculated EWSR with the data of Ref. \cite{HartmannBabilonKamerdzhievEtAl2004}, at that we show slightly different endpoint energies for the cases, where substantial strength is concentrated around 10 MeV ($^{48}$Ca) or where the particle emission threshold is below 10 MeV ($^{40}$Ca). Thus, within the theoretical and experimental error bars the agreement is very reasonable. 

\section{Summary}
\label{summary}

New results for dipole strength distributions in stable and unstable neutron-rich Calcium isotopes are presented. The calculations were performed within the relativistic quasiparticle time blocking approximation including effects of coupling between quasiparticles and phonons. Thus, (i) the performance of RQTBA has been tested in the low mass region and (ii) the systematics of the pygmy dipole resonance has been extended to neutron rich Calcium isotopes beyond the closed-shell $^{48}$Ca. In general, the obtained results for both giant dipole resonance and the low-energy fraction of the dipole strength distribution are very satisfactory. The case of $(Z,N)$ = (20,20) shells closures requires, however, a special consideration because the low-lying phonon modes with positive nature parity can not be reproduced in the standard R(Q)RPA framework and require an extended approach for the phonons.

Besides this, some other issues with the dipole response in Calcium mass region remain unsolved.
In the context of the previous R(Q)TBA calculations of the dipole strength \cite{LitvinovaRingTselyaev2007}, one can see that there is a common problem for the models based on the meson-exchange Lagrangian adjusted in the standard way to bulk nuclear properties: the GDR centroids are systematically lower at the $(Z,N)$ = (20,20), $(Z,N)$ = (50,82) and $(Z,N)$ = (82,126) double shell closures. 
This problem appears on the CDFT (R(Q)RPA) level, and the inclusion of particle-vibration coupling in the conventional resonant time blocking approximation does not shift the centroid, thus this problem persists in RQTBA. However, before addressing this issue on the CDFT level, one might still consider going beyond the conventional quasiparticle-phonon coupling scheme of RQTBA, which includes only isoscalar normal phonons. For instance, coupling to pairing vibrations can be of special importance at the shell closures and coupling to isovector phonons can potentially play a role as they represent a dynamical approach to proton-neutron pairing \cite{Litvinova2016}. These issues are under consideration and will be studied in future endeavors.

%
\section*{Acknowledgements}
%
%
This work is partly supported by US-NSF grants PHY-1204486 and PHY-1404343. I.A.E. also acknowledges support from FAIR-Russia Research Center.


\bibliography{Ca-thNotes1}
\end{document}